\documentclass[floatfix,aps,amsmath,prd,twocolumn,notitlepage,eqsecnum,showpacs,nofootinbib,aasmacros,superscriptaddress]{revtex4-1}
\usepackage{bm, graphics, graphicx, epsfig, soul, latexsym, hyperref ,multirow}
\usepackage{hyperref}
\usepackage{comment}
\usepackage{subfigure}
\usepackage{graphicx,color}
\usepackage{dcolumn}
\usepackage{bm,amsmath, amssymb}
\usepackage{mathrsfs}
\usepackage{bigints}
\usepackage{float}
\usepackage{multirow}
\usepackage{footmisc}
\usepackage{placeins}
\usepackage[normalem]{ulem}

\usepackage{amsmath}

\begin{document}


\title{Premerger localization of intermediate mass binary black holes with LISA and prospects of joint observations with Athena and LSST}

\author{Pankaj Saini}
\email{pankajsaini@cmi.ac.in}
\affiliation{Chennai Mathematical Institute, Plot H1 SIPCOT IT Park, Siruseri 603103, India.}
\author{Sajad A. Bhat}
\email{sabhat@cmi.ac.in}
\affiliation{Chennai Mathematical Institute, Plot H1 SIPCOT IT Park, Siruseri 603103, India.}
\author{K. G. Arun}
\email{kgarun@cmi.ac.in}
\affiliation{Chennai Mathematical Institute, Plot H1 SIPCOT IT Park, Siruseri 603103, India.}
\date{\today}
\begin{abstract}
The planned Laser Interferometric Space Antenna (LISA) will be able to detect gravitational waves (GWs) from intermediate mass binary black holes (IMBBHs) in the mass range $\sim 10^{2} \mbox{--} 10^{4}\,M_{\odot}$ up to a redshift $z\sim20$. Modulation effects due to LISA’s orbital motion around the Sun facilitate precise premerger localization of the sources, which in turn would help in electromagnetic (EM) followups. In this work, we calculate the uncertainties in sky position, luminosity distance, and time of coalescence as a function of time to coalescence. For representative masses of the IMBBHs, we synthesize a population of binaries uniformly located and oriented on a sphere of radius 3 Gpc and compute the projected parameter measurement uncertainties using the Fisher information matrix. We find that for systems with a total mass of $10^3\,M_{\odot}$, the errors in the sky position and luminosity distance are $\sim 0.4\,\text{deg}^2$ and $\sim 6\%$, respectively, one day prior to coalescence. The coalescence time can be predicted with an uncertainty $\lesssim 10$ sec, one day before coalescence. We also find that for $10^3\,M_{\odot}$, around $40\%$ ($100\%$) of the population has a source localization that is smaller than the field of view of Athena (LSST) one day before the merger. These extremely precise measurements can be used to alert ground-based GW detectors and EM telescopes about the time and location of these mergers. We also discuss mechanisms that may produce EM emission from IMBBH mergers and study its detectability using the planned Legacy Survey of Space and Time (LSST) in the optical and Athena in the x-ray bands. Detection of an EM transient may provide us vital clues about the environments where these mergers occur and the distance estimation can pave the way for cosmography.
\end{abstract}

\maketitle

\section{Introduction}
Existence of intermediate mass black holes (IMBHs) with masses in the range $10^2\mbox{--}10^4\,M_{\odot}$ has been a long-standing puzzle in astronomy (see Refs.~\cite{Miller:2003sc,greene2020intermediate} for reviews). Galactic x-ray binaries~\cite{BlackCat} and gravitational wave observations by LIGO/Virgo~\cite{GW150914,GWTC-3} have established the presence of stellar mass black holes with masses of tens of solar masses. There are several compelling evidences for the existence of supermassive black holes, with masses ranging from millions to billions of solar masses, in various bands of the electromagnetic (EM) spectrum (see for instance Refs.~\cite{Kormendy:1995er,Kormendy:2013dxa,Ghez:1998ph,Schodel:2002py, Genzel2010}) including the most recent observations by the Event Horizon Telescope~\cite{M87-EHT,MW-EHT}. IMBHs present the missing link between the stellar mass and supermassive black holes.

Since stellar processes are unlikely to produce black holes (BHs) of mass $\gtrsim 60 \,M_{\odot}$, heavier BHs are likely to be produced by the repeated mergers of stars and/or stellar remnants inside dense stellar environments such as young clusters, globular clusters, nuclear star clusters and galactic nuclei ~\cite{PhysRevD.100.043027,giersz2015mocca,Bellovary:2015ifg,Stone:2016ryd,Rose:2021ftz,Fragione:2021nhb,Mapelli:2021syv,Gonzalez:2020xah,DiCarlo:2021att}. Merger rates of intermediate mass binary black holes (IMBBHs) in nuclear star cluster are expected to be $\sim 0.01$-$10 \,\text{Gpc}^{-3}\text{yr}^{-1}$ depending on their mass~\cite{Fragione:2022avp}. IMBHs can also be formed through the direct collapse of a gas cloud at high redshifts. This channel can lead to the formation of $10^4 \mbox{-}10^5 \,M_{\odot}$ BHs \cite{Begelman:2006db}. The centers of dwarf galaxies are also expected to be the potential sites for hosting IMBHs~\cite{baldassare201550,2021ApJ...920..134P}.

However, the observational evidences for IMBHs using EM observations are not as compelling as for the other two classes. Although there are several candidates, none of them is considered to be conclusive evidence for IMBHs \cite{2009Natur.460...73F,Pasham:2014ybe,2017Natur.542..203K,Mezcua:2017npy}. (See \cite{greene2020intermediate} for a review of different types of IMBH searches.) More recently, gravitational wave (GW) observations using LIGO~\cite{LIGOScientific:2014pky} and Virgo~\cite{VIRGO:2014yos} have emerged as a new tool for probing IMBHs. In the third GW transient catalog (GWTC-3)~\cite{GWTC-3}, three binaries have total mass above $100 \,M_{\odot}$~\footnote{These are systems for which even the lower mass end of the posterior distribution at $90\%$ credibility, lie above $100\,M_{\odot}$.}, the heaviest one being $\sim\, 150\, M_{\odot}$~\cite{GW190521}. This, though at the lower end of the IMBH mass spectrum, provides the cleanest evidence for their existence and further motivates extensive searches for them. 

While ground-based detectors may not be able to see such mergers beyond a few thousands of solar masses, future space-based detector Laser Interferometric Space Antenna (LISA)~\cite{Gair:2017ynp,amaro2017laser} will be capable of detecting coalescing binary black holes (BBHs) in the total mass range $\sim 10^3\mbox{-}10^7 \,M_{\odot}$ up to and beyond a redshift of $z \sim 20$ \cite{amaro2017laser,Jani:2019ffg}. More precisely, binaries with a total mass of around $10^5\, M_{\odot}$ will inspiral and merge within the LISA sensitivity band ($10^{-4}\mbox{--}0.1$ Hz). Less massive binaries with total mass $\sim 10^4 \, M_{\odot}$ will inspiral in LISA's frequency band, but merge outside the frequency band of LISA. Lower mass BHs ($\lesssim 10^3 \,M_{\odot}$) will still inspiral in the LISA band but merge in the frequency band of ground-based detectors and hence, facilitate multiband observations of these binaries ~\cite{Vitale:2016rfr, Sesana:2016ljz, Cutler:2019krq}.

LISA has a constellation of three spacecraft arranged in the form of an equilateral triangle with each side having a length of 2.5 million kilometers~\cite{amaro2017laser}. This constellation rotates around its own axis and orbits around the Sun with a period of one year. These motions of LISA lead to modulations to the amplitude and phase of the GW signal which encodes information about the sky position and orientation of the source~\cite{Wahlquist:1987rx,Cutler:1997ta}. Therefore, LISA will be able to locate the source position with high precision. The sources which spend larger number of GW cycles in the band can be localized with better precision due to larger number of modulations in the amplitude and GW phase~\cite{Cutler:1997ta,Arun:2007hu,Trias:2007fp}.
 
If an IMBBH merger happens in a gaseous environment such as in an active galactic nucleus (AGN) disc, it is possible that the interaction of the merger remnant with the ambient gas-rich medium can lead to accretion onto the BH thereby producing EM flares~\cite{Kaaret:2017tcn,McKernan:2019hqs}. LISA can guide the EM searches for any associated transients with optical, radio, and x-ray wide-field instruments, such as LSST \cite{abell2009lsst}, SKA~\cite{2009SKA}, and Athena \cite{McGee:2018qwb,Piro:2021oaa}. Due to the high signal-to-noise ratio (SNR) attained near the coalescence, sky resolution is expected to improve to few arcminutes close to the merger which might help in exploring the features of postmerger dynamics by deeper EM searches~\cite{Babak:2008bu}. Moreover, the joint EM and GW detections of IMBBHs will provide a special class of standard sirens \cite{tamanini2016science}, thus probing the history of cosmic expansion.

There have been several recent works that investigated the parameter estimation of stellar mass~\cite{toubiana2020parameter}, intermediate mass~\cite{Caputo:2020irr} and supermassive~\cite{mangiagli2020observing} binary black holes focusing on the improvements due to the inclusion of higher modes, spin-induced precession, and orbital eccentricity (see, for example, Refs.~\cite{Arun:2007hu, Trias:2007fp,Babak:2008bu, Arun:2008zn,Stavridis:2009ys,Klein:2009gza} for similar works in the context of the old LISA configuration.). The inclusion of new features in the waveform leads to improved parameter estimation in general. Effects of multiband observations of stellar mass binary black holes on the source localizability have also been studied in various works~\cite{buscicchio2021bayesian} (see~\cite{Muttoni:2021veo,Zhu:2021bpp} for implications in the multibanding in the context of cosmology). The effect of a future network of space-based detectors on the source localization was also studied in Ref.~\cite{Zhang:2021wwd}.

References \cite{colpi2019athena,Piro:2021oaa} discussed the synergy between LISA and EM observations in detail. Reference \cite{McGee:2018qwb} studied the detection of x-ray counterpart from massive and stellar mass black hole binaries with Athena given the early detection with LISA.  More recently several works have also investigated the detection of EM counterpart associated with LISA sources, especially in the radio, optical, and x-rays~\cite{Mangiagli:2022niy}. 

In this work, we discuss projected parameter measurement errors of IMBBHs ($\sim$ $500-10^{4}\, M_{\odot}$) in the LISA band at different times prior to coalescence and its importance in the context of astrophysics especially focusing on uncertainty in source localization and distance estimation. For each of the representative systems considered, we synthesize a population of $10^3$ sources at a luminosity distance of 3 Gpc with the sky position and orientation of them uniform on the surface of the corresponding sphere. Using Fisher matrix analysis, we calculate parameter measurement errors for these $10^3$ sources as a function of time before merger. We find that for the system with total mass $10^3 \, M_{\odot}$, the median accuracy on angular resolution and luminosity distance are $\sim 0.4\, \text{deg}^2$ and $\sim 6\%$, respectively one day prior to the coalescence. Achieving this level of accuracy prior to merger provides a platform for providing early warning for EM observatories in search of a potential EM counterpart associated with the merger. We also discuss, on general grounds, some of the mechanisms that will generate EM counterparts associated with these mergers in x-ray and optical bands and discuss their detectability with Athena and LSST, respectively.

The rest of the paper is organized as follows: In Sec. \ref{waveform-model}, we describe the waveform model without averaging over the sky position and binary orientations and discuss how modulational effects arise due to the motion of LISA. Section \ref{parameter estimation} explains the Fisher matrix formalism for calculating the statistical errors on binary parameters. In Sec. \ref{results} we discuss our findings and results. Section \ref{implications} presents the astrophysical implications from optical and x-ray observations. In Sec. \ref{conclude}, we summarize and conclude our discussions. Throughout the paper we use geometric units ($G=c=1$).

\section{Waveform model and Parameter uncertainty estimation}
The waveform model for compact binaries taking LISA motion into account has been developed in Ref.~\cite{Cutler:1997ta}. We briefly summarize the essential ingredients of the waveform model and the projected parameter measurement uncertainty for LISA closely following \cite{Cutler:1997ta,berti2005estimating}.
\subsection{Waveform model}\label{waveform-model}

LISA can be considered as a combination of two L-shaped detectors due to its triangular shape. The strain $h(t)$ produced by GWs in the detector is given as 
\begin{equation}\label{waveform}
    h_{\rm \alpha}(t) = \frac{\sqrt{3}}{2} \bigg(F^{+}_{\rm \alpha}(t)h_{+}(t) + F^{\times}_{\rm \alpha}(t)h_{\times}(t)\bigg),
\end{equation}
where $\alpha=\rm I,II$ represents the first and the second detector. The factor $\sqrt{3}/2$ is due to the $60^{\circ}$ angle between adjacent arms of LISA. $F_{\rm \alpha}^{+,\times}$ are the detector antenna pattern functions which depend on the sky position $(\theta_{S},\phi_{S})$ of the source and the polarization angle $(\psi_{S})$. The unbarred angles ($\theta_{S},\phi_{S},\psi_{S}$) are defined in the rotating LISA-centric coordinate system which changes with time as the detector moves. These angles are reexpressed in terms of the barred angles ($\bar{\theta}_{S}, \bar{\phi}_{S}, \bar{\theta}_{L}, \bar{\phi}_{L})$ which fix the position and orientation of the source with respect to the solar barycenter frame. Here the subscript ``$S$'' stands for the source and ``$L$'' stands for the orbital angular momentum of the binary. Detailed discussion about antenna pattern functions and the expressions relating barred and unbarred angles can be found in Refs. \cite{Cutler:1997ta,berti2005estimating}. 

The amplitudes of $+$ and $\times$ polarizations, $h_{+}(t)$ and $h_{\times}(t)$, in Eq.~\eqref{waveform} are given as
\begin{equation}
    h_{+}(t) = A(t) (1+({\bf\hat{\hspace{0.5mm}L}} \cdot {\bf\hat{ n}})^{2}\big) \, ,
\end{equation}
\begin{equation}
    h_{\times}(t) = -2 A(t) ({\bf\hat{\hspace{0.5mm}L}} \cdot {\bf\hat{n}}\big)\, .
\end{equation}
Here 
\begin{equation}
    A(t) = \frac{2 m_1 m_2}{r(t) D_L}\, ,
\end{equation}
where $m_1$ and $m_2$ are the component masses of binary, $r(t)$ is the separation between two bodies, and $D_L$ is the luminosity distance to the source, $\bf{\hat{\hspace{0.5mm}L}}$ and $-\bf{\hat{n}}$ are the unit vectors along the directions of orbital angular momentum and GW propagation, respectively.

Taking LISA's orbital motion into account, Eq.~\eqref{waveform} can be rewritten in terms of the amplitude and phase as 
\begin{equation}\label{fullwaveform}
h_{\rm \alpha}(t) = \frac{\sqrt{3}}{2} A(t) A_{p,\rm \alpha}(t) \cos\Big[\int_{0}^{t} f(t')dt' + \phi_{p,\rm \alpha}(t) + \phi_{D}(t) \Big] \, .
\end{equation}
The polarization amplitude $A_{p,\rm \alpha}(t)$, polarization phase $\phi_{p,\rm \alpha}(t)$, and Doppler phase $\phi_{D}(t)$ are given as \cite{Cutler:1997ta,berti2005estimating}

\begin{equation}
  A_{p,\rm \alpha}(t) = \sqrt{F_{\rm \alpha}^{+}(t)^{2} \big(1+({\bf\hat{\hspace{0.5mm}L}} \cdot {\bf{\hat n}})^{2}\big)^{2} + 4 F_{\rm \alpha}^{\times}(t)^{2}\big({\bf\hat{\hspace{0.5mm}L}} \cdot {\bf{\hat n}}\big)^{2} \big)} \, ,
\end{equation}
\begin{equation}\label{polarisation}
\phi_{p,\rm \alpha}(t) = \tan^{-1}\bigg(\frac{2 F_{\rm \alpha}^{\times}(t)({\bf\hat{\hspace{0.5mm}L}}\cdot {\bf\hat{ n}})}{F_{\rm \alpha}^{+}(t)(1+({\bf\hat{\hspace{0.5mm}L}}\cdot {\bf{\hat n}})^{2})}\bigg) \,,
\end{equation}
\begin{equation}\label{doppler}
    \phi_{D}(t) = 2 \pi f(t) R \sin\bar{\theta}_{S}  \cos\big[\bar{\phi}(t)-\bar{\phi}_{S} \big]\, ,
\end{equation}
here $R=1$ AU is the distance between Earth and Sun, $\bar{\phi}(t) = \bar{\phi}_{0}+2\pi t/T$, where $T=1$ year is the orbital period of LISA spacecraft around the Sun, $\bar{\phi}_{0}$ is a constant which specifies the detector location at time $t=0$. We choose $\bar{\phi}_{0} =0$.

Note that $A_{\rm \alpha}(t)$, $\phi_{p,\rm \alpha}(t)$, and $\phi_{D}(t)$ change on timescales of the order of one year which is much larger than the orbital period of binary, so we can write Eq.~\eqref{fullwaveform} in the frequency domain using stationary phase approximation (SPA) as~\cite{berti2005estimating}
\begin{align}\label{fourierwave}
\Tilde{h}_{\rm \alpha}(f) = & \frac{\sqrt{3}}{2} \mathcal{A} f^{-7/6} e^{i \Psi (f)} \bigg[\frac{5}{4} A_{p,\rm \alpha}(t(f))\bigg] \nonumber\\ 
& e^{-i \big(\phi_{p,\rm \alpha}(t(f)) + \phi_{D}(t(f))\big)}.
\end{align}
Here
\begin{equation}
   \mathcal{A}= \frac{1}{\sqrt{30}}\frac{\mathcal{M}^{5/6}}{\pi^{2/3}D_{L}} \, ,
\end{equation}
where $\mathcal{M}$ = $\eta^{3/5} M$ is the chirp mass of the system, $\eta=(m_1 m_2)/(m_1+m_2)^{2}$ is the symmetric mass ratio and $M=m_1+m_2$ is the total mass of the system and $D_L$ is the luminosity distance to the source. Note that $M$ here and throughout is the source frame mass of the binary. We multiply by a factor of $(1+z)$ to convert the source frame mass to the detector frame mass, $ M_{\rm obs} \xrightarrow{} (1+z) M$, where $z$ is the cosmological redshift of the source. We use the redshift-distance relation for a flat universe \cite{hogg} as
\begin{equation}
\label{eq:dLz}
D_L = \frac{(1+z)}{H_0} \int_0^z \frac{dz'}{\sqrt{\Omega_M (1+z')^3 + \Omega_{\Lambda}}},
\end{equation}
where the cosmological parameters are given in Ref.~\cite{Planck:2015fie} as:
{$H_{0}=67.90$(km/s)/Mpc}, $\Omega_{M}=0.3065$, and $\Omega_{\Lambda}=0.6935$.

We use the frequency-domain {\tt TaylorF2} waveform approximant in our calculation. The waveform model describes the inspiral part of the GW signal and assumes that spins are aligned or antialigned with the orbital angular momentum of the binary. Since we are interested in the extrinsic parameters of the IMBBHs which inspiral in the LISA band and merge outside the LISA band, the use of {\tt TaylorF2} waveform is sufficient for our purpose. 

In post-Newtonian (PN) theory~\cite{Blanchet:1995ez,Blanchet:1995fg,Kidder:1995zr,Blanchet:2002av,Blanchet:2006gy,BDEI04,DIS3,DISupdate,Arun:2008kb,Marsat:2012fn,Mishra:2016whh}, the SPA phase $\Psi(f)$ in Eq.~\eqref{fourierwave} can be written as a  power series in orbital velocity $v=(\pi M f)^{1/3}$ as
\begin{equation}
\label{GWphase}
     \Psi(f)  =  2\pi ft_{c} + \phi_{c} + \frac{3}{128 \eta v^{5}} \sum_{k} (\phi_{k}v^{k} 
      +  \phi_{kl}  v^{k} \ln v) \,,
\end{equation}
where $t_{c}$ and $\phi_{c}$ are the time and phase of coalescence. The coefficients $\phi_{k}$ are called the PN coefficients which are the function of intrinsic parameters of the source such as mass and spin. The coefficient with $v^{k}$ term relative to the leading order is referred to as the ($k/2$)PN coefficient. The full expressions for $\phi_k$ and $\phi_{kl}$ up to 3.5PN order can be found in Refs.~\cite{Arun:2004hn,Arun:2006hn,Arun:2008kb,Buonanno:2009zt,Mishra:2016whh}. Using the energy-balance equation, adiabatic approximation and Kepler's law, the expression for $t(v)$ or equivalently $t(f)$ in Eq.~\eqref{fourierwave} can be computed as 
\begin{equation}
    t(v) = t_{\rm  ref} +  \int^{v_{\rm  ref}}_{v} dv\frac{E^{'}(v)}{\mathcal{F}(v)}\,,
\end{equation}
where $t_{\rm ref}$ is an integration constant, $v_{\rm ref}$ is an arbitrary reference velocity, $E^{'}(v)= dE(v)/dv$, where $E(v)$ is the binding energy of the system and $\mathcal{F}(v)$ is the gravitational wave flux. The full expressions for $E(v)$ and $\mathcal{F}(v)$ can be found in Refs.~\cite{Buonanno:2009zt,Arun:2008kb,Arun:2009mc}.

\subsection{Projected parameter measurement uncertainty using the Fisher matrix}\label{parameter estimation}
We use Fisher information matrix framework~\cite{Finn:1992wt,Cutler_Flanagan,Clifford_Will} to calculate the statistical errors on binary parameters. In the limit of large SNR, stationary and Gaussian noise, this framework gives the projected 1$\sigma$ width of the posterior probability distribution on binary parameters. (see Ref.~\cite{vallisneri2008use,Rodriguez:2013mla} for some caveats associated with the use of this framework for science case studies.) The projected posterior probability distribution (under the assumption of stationary, Gaussian noise and large SNR) on binary parameters $\bm \theta$ given the detector output $s(t)$ can be approximately written as 
\begin{equation}\label{posterior}
 p({\bm \theta}|s) \propto p^{0}({\bm \theta}) \exp\left[ -\frac{1}{2} \Gamma_{ab} (\theta^{a} - \hat{\theta}^{a}) (\theta^{b} - \hat{\theta}^{b}) \right]\,, \end{equation}
where $p^{0}({\bm \theta})$ is the prior distribution about the parameters of the signal characterized by $\bm\theta$. The values $\hat{\theta}^{a}$ are the ``true'' parameters that maximize the likelihood. In Eq.~\eqref{posterior}, $\Gamma_{ab}$ is the Fisher information matrix. Since LISA can be considered as a combination of two L-shaped detectors $\rm I$ and $\rm II$, the total Fisher matrix can be written as
\begin{equation}\label{combined fisher}
     \Gamma_{ab} = \bigg(\frac{\partial{h_{\rm I}}}{\partial\theta^a}\bigg|\frac{\partial{h_{\rm I}}}{\partial\theta^b}\bigg) + \bigg(\frac{\partial{h_{\rm II}}}{\partial\theta^a}\bigg|\frac{\partial{h_{\rm II}}}{\partial\theta^b}\bigg)\, ,
\end{equation}
where Eq.~\eqref{combined fisher} is evaluated at the measured value $\hat{\theta}^{a}$ of the parameters $\bm \theta$. In Eq.~\eqref{combined fisher} $(|)$ denotes the noise weighted inner product which for the two signals $a(t)$ and $b(t)$ is defined as
\begin{equation}\label{inner product}
(a|b)= 2 \int^{f_{\rm high}}_{f_{\rm low}}df\,\frac{\tilde a^{*}(f)\,\tilde b(f)+\tilde a(f)\, 
\tilde b^{*}(f)}{S_n(f)}\, ,
\end{equation}
where $\tilde{a}(f)$ ($\tilde{b}(f)$) is the Fourier transform of $a(t)$ ($b(t)$) and $\ast$ represents the complex conjugation and $S_n(f)$ is the one-sided noise power spectral density (PSD) of the detector. The SNR ($\rho$) is defined as 
\begin{equation}\label{snr}
    \rho^2 = (h_{\rm I}|h_{\rm I}) + (h_{\rm II}|h_{\rm II})= \rho_{\rm I}^2 + \rho_{\rm II}^2\, .
\end{equation}
We assume that our prior distribution of the model parameters corresponds to a Gaussian distribution centered around $\bar{\theta}^a$,
\begin{equation}
 p^0({\bm \theta}) \propto \exp\bigg[ -\frac{1}{2} \Gamma^0_{ab} (\theta^a-\bar{\theta}^a) (\theta^b-\bar{\theta}^b) \bigg],
\end{equation}
where $\Gamma^0_{ab}$ is the prior matrix. Assuming $\bar{\theta}^a \approx \hat{\theta}^a$, the covariance matrix which refers to the covariance of the posterior distribution under the Gaussian approximation,
is given as
\begin{equation}\label{covariance}
\Sigma_{ab} = (\Gamma_{ab}+\Gamma_{ab}^0)^{-1} \,. 
\end{equation}
The 1$\sigma$ statistical errors in binary parameters $\theta^a$ are given as
\begin{equation}\label{error}
    \sigma_a = \sqrt{\Sigma_{aa}}.
\end{equation}

Our parameter space consists of the following parameters,
\begin{equation}
 \theta^{a} = \{t_{c}, \phi_{c}, \mathcal{M}, \eta, D_{L}, \bar{\theta}_{S}, \bar{\phi}_{S},\bar{\theta}_{L}, \bar{\phi}_{L},\chi_{1},\chi_{2}\} \,,
\end{equation}
where $\chi_1$ and $\chi_2$ are the dimensionless spins of component BHs.
The physically allowed values of coalescence phase $\phi_{c}$ and spin parameters ($ \chi_{1}$,$\chi_{2}$) are restricted to the ranges $\phi_{c} \in [-\pi, \pi]$ and $\chi_{1,2} \in [-1, 1]$, respectively. This is taken into account by adopting Gaussian priors on $\phi_c$ and $\chi_{1,2}$ with zero means and 1$\sigma$ widths. The 1$\sigma$ width on $\phi_c$ and $\chi_{1,2}$ is given by $\delta \phi_c =\pi$ and $\delta \chi_{1,2}=1$, respectively. To incorporate these priors, a prior matrix with the nonzero components, $\Gamma^0_{\rm \phi_c,\phi_c}=1/\pi^2$, $\Gamma^0_{\rm \chi_1,\chi_1}=\Gamma^0_{\rm \chi_2,\chi_2}=1$ is added to the Fisher matrix.

\begin{figure}[h]
    \centering
    \includegraphics[width=0.94\linewidth]{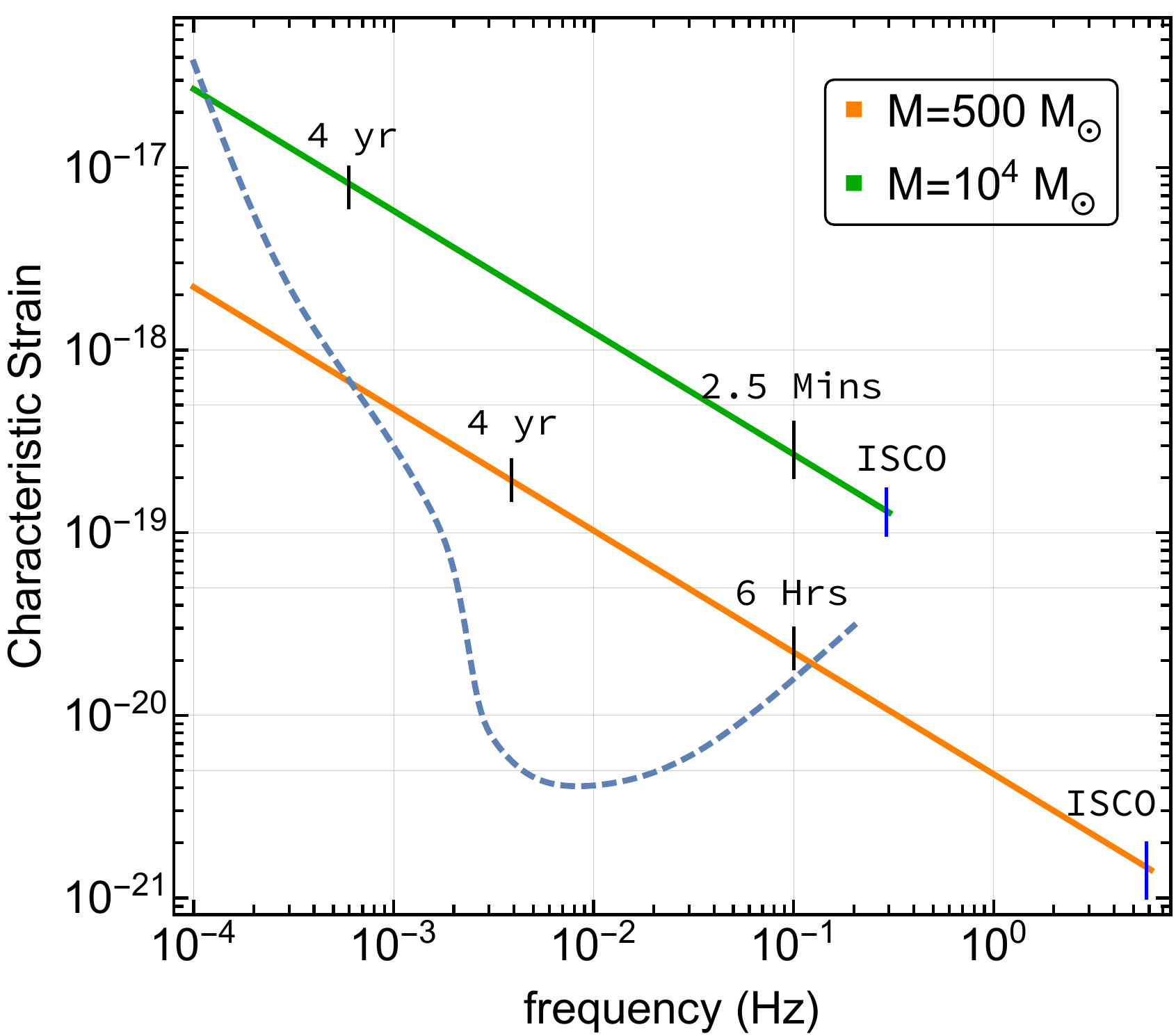}
    \caption{(Color online) The noise power spectral density of LISA (dashed curve) and the evolution of two representative IMBBH systems in the LISA frequency band (solid lines). We assume mass ratio for the sources to be $m_1/m_2=2:1$. The luminosity distance is fixed to be $3$ Gpc. The markers in blue show the characteristic strain at ISCO. The black markers represent the time remaining before reaching ISCO.}
    \label{noise LISA}
\end{figure}

The noise sensitivity curve of LISA consists of instrumental noise and white dwarf confusion noise. The instrumental noise PSD is taken from Eq.~(1) of Ref.~\cite{babak2017science}. The galactic confusion noise can be found in Eq.~(4) of Ref.~\cite{mangiagli2020observing}. Note that the instrumental noise in Ref.~\cite{babak2017science} is sky-averaged and accounts for the $60^{\circ}$ angle due to the triangular shape of LISA. Since we take LISA motion into account, we use a non-sky-averaged noise PSD. As we include a factor $\sqrt{3}/2$ in our definition of GW signal, we multiply by a factor of $3/20$ in Eq.~(1) of Ref.~\cite{babak2017science} to obtain a non-sky-averaged noise PSD. The lower and upper cutoff frequencies ($f_{\rm low}$ and $f_{\rm high}$) in Eq.~\eqref{inner product} are fixed by the sensitivity of the detector, observation time, and properties of the source. The $f_{\rm low}$ is given by
\begin{equation}\label{flow}
f_{\rm low}={\rm max}  \{10^{-4},f_{\rm year}\} \,.
\end{equation}
Here, $f_{\rm year}$ is given as~\cite{berti2005estimating}
\begin{equation}
f_{\rm year} = 4.149\times 10^{-5}\left(\frac{{\cal M}}{10^6 M_\odot}\right)^{-5/8}
\left(\frac{T_{\rm obs}}{1 \rm \text{year}}\right)^{-3/8} \, ,   
\end{equation}
where $T_{\rm obs}$ is the observation time before the innermost stable circular orbit (ISCO). In our calculation, we assume $T_{\rm obs}=4$ year\footnote{Note that IMBBHs we consider will not spend exactly $4$ year in the LISA band, since they exit the LISA frequency band at $0.1$ Hz and merge at higher frequencies.}. The upper cut-off frequency is given as
\begin{equation}
    f_{\rm high} = {\rm min}\big\{0.1,f_{\rm ISCO}\big\}\,,
\end{equation}
where $f_{\rm ISCO}$ is the GW frequency corresponding to the Schwarzschild ISCO and reads as
\begin{equation}
    f_{\rm ISCO} = \frac{1}{(6^{3/2}\pi M)}\,.
\end{equation}
In Fig.~\ref{noise LISA}, we show the noise curve for LISA. We also show the characteristic strain (pattern and inclination angle averaged) for two representative IMBBH sources with total masses $M=500\, M_{\odot}$ and $M=10^4\, M_{\odot}$ at fixed mass ratio $m_1/m_2=2:1$. The luminosity distance to the sources is fixed to be $3$ Gpc. The vertical black lines mark the times before reaching ISCO. These systems sweep through LISA's frequency band and merge outside the LISA band after exiting at $0.1$ Hz.

The uncertainty in the sky position (angular resolution) of the source is defined as \cite{Cutler:1997ta,berti2005estimating}
\begin{equation}
    \Delta\Omega_{S} = 2 \pi |\sin\bar{\theta}_{S}| \bigg(\Sigma^{\bar{\theta}_{S}\bar{\theta}_{S}}\Sigma^{\bar{\phi}_{S}\bar{\phi}_{S}} - (\Sigma^{\bar{\theta}_{S}\bar{\phi}_{S}})^2\bigg)^{1/2},
\end{equation}
where $\Sigma$ is the covariance matrix defined in Eq.~\eqref{covariance}. 
Similarly, the polarization resolution can be defined as 
\begin{equation}
    \Delta\Omega_{L} = 2 \pi |\sin\bar{\theta}_{L}| \bigg(\Sigma^{\bar{\theta}_{L}\bar{\theta}_{L}}\Sigma^{\bar{\phi}_{L}\bar{\phi}_{L}} - (\Sigma^{\bar{\theta}_{L}\bar{\phi}_{L}})^2\bigg)^{1/2}.
\end{equation}
Polarization resolution, in our context, refers to the ability to accurately determine the direction of angular momentum of the binary~\cite{Jennrich:1997if}. For example, if there is a jetted electromagnetic emission following the merger, similar to the case of a short gamma-ray burst following a binary neutron star merger, $\Delta \Omega_L$ would provide the accuracy with which such a jet can be resolved~\cite{Arun:2014ysa}.

\section{Results}\label{results}
Using the waveform model discussed in Sec.~\ref{waveform-model} and the Fisher matrix formalism discussed in Sec.~\ref{parameter estimation}, we calculate SNR and 1$\sigma$ statistical errors on binary parameters at different times prior to coalescence. We consider four representative IMBBH systems with total masses $M= (500,\, 10^3,\, 5\times10^3,\, 10^4)\,M_{\odot}$. The mass ratio for all the sources is fixed to be $m_1/m_2=2:1$. The dimensionless spin parameters are assumed to be aligned with the orbital angular momentum of the binary and their magnitudes $(\chi_{1}, \chi_{2}$) are chosen to be $0.5$ and $0.4$, respectively. The choice of spin parameters is arbitrary and does not alter our conclusions, as we are focusing on the measurement of extrinsic parameters of the binary which are known to be relatively uncorrelated with the spin parameters~\cite{Kocsis:2007hq}. We assume that all the systems are fixed at $D_L=3$ Gpc. This choice of distance is made keeping in mind LISA's SNR threshold ($\rho_{\rm th}=10$) and expected merger rates of IMBBHs ($\sim 0.01\mbox{--}10 \,   \text{Gpc}^{-3}\text{yr}^{-1}$ \cite{Fragione:2022avp}). Hence, choosing $D_L=3$ Gpc is a trade-off between SNR and IMBBH merger rates and ensures, with our limited knowledge of the IMBBH population, that there is a reasonable chance of detecting the systems studied here.

As the errors depend crucially on the location and orientation of the source, we synthesize a population for each of the representative systems considered and use the median value of the resulting distribution to assess the parameter measurement uncertainty~\cite{berti2005estimating}. Towards this, for each of these binary systems we distribute $10^3$ sources uniformly over sky position and orientation. The positions and orientations of these binaries in the heliocentric orbit's frame, $(\bar{\theta}_{S}, \bar{\phi}_{S})$ and $(\bar{\theta}_{L}, \bar{\phi}_{L})$ are randomly sampled from a uniform distribution over a sphere. In other words, the angles $\bar{\phi}_{S}$ and $\bar{\phi}_{L}$ are randomly generated from the uniform distribution [$0,2\pi$] and $\cos\bar{\theta}_{S}$, $\cos\bar{\theta}_{L}$ are drawn in the range [$-1,1$]. For each source, we calculate the Fisher matrix according to Eq.~\eqref{combined fisher} at 1 month, 1 week, 1 day, and 1 hour before their coalescence.

\begin{figure}[h]
    \centering
    \includegraphics[width=0.96\linewidth]{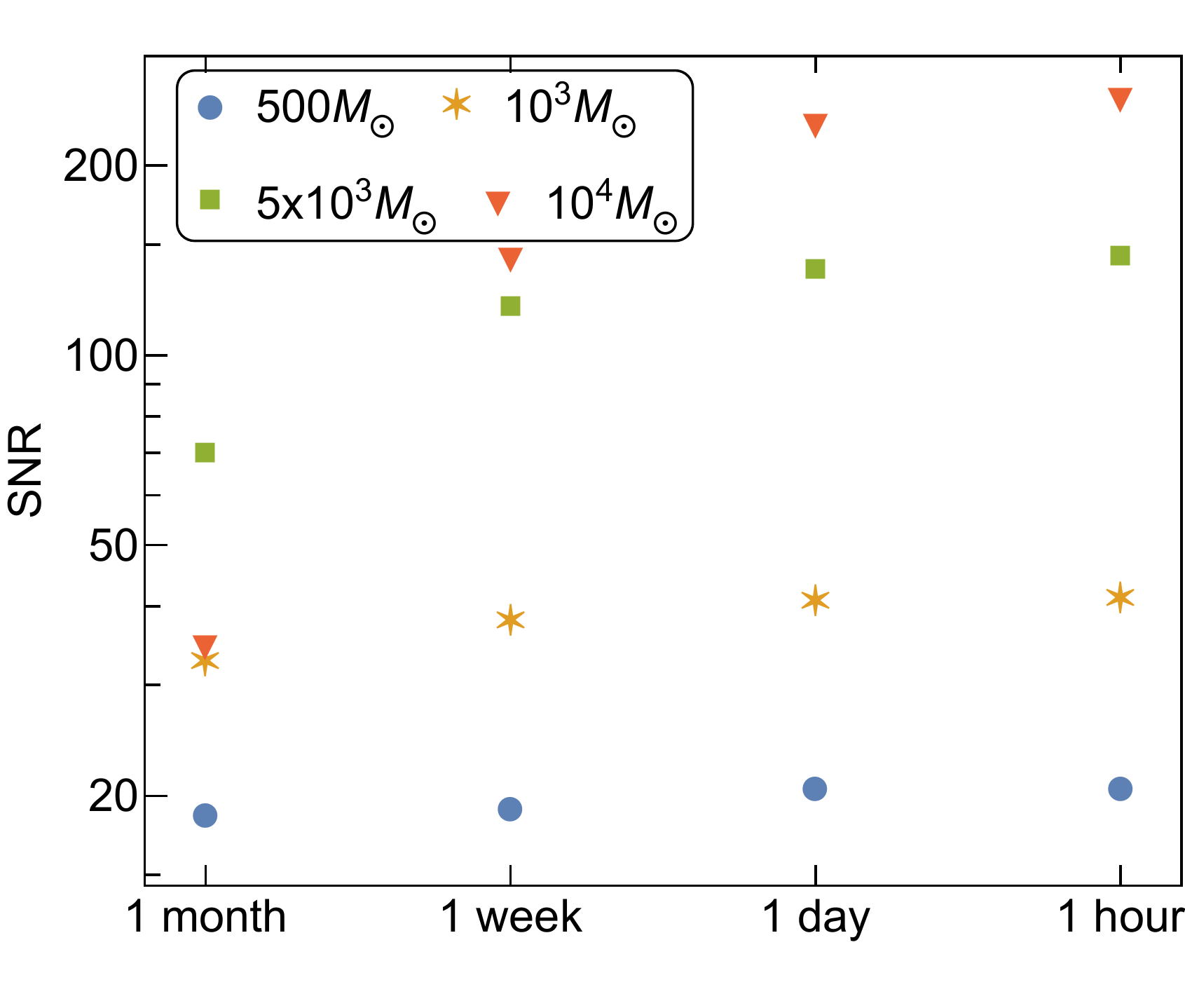}
    \caption{(Color online) Accumulated SNR as a function of time to coalescence for four representative systems. For all systems the mass ratio ($m_1/m_2$) is fixed to be $2:1$ and the dimensionless spin parameters are $\chi_1=0.5$, $\chi_2=0.4$. All the systems are located at a luminosity distance of $3$ Gpc. For each system, $10^3$ realizations are distributed over randomly sampled sky position and orientation from a uniform distribution. Median values from these realizations are plotted.}
    \label{snr}
\end{figure}

\begin{figure*}[tph]
    \centering 
    \begin{subfigure}{\includegraphics[width=0.48\linewidth]{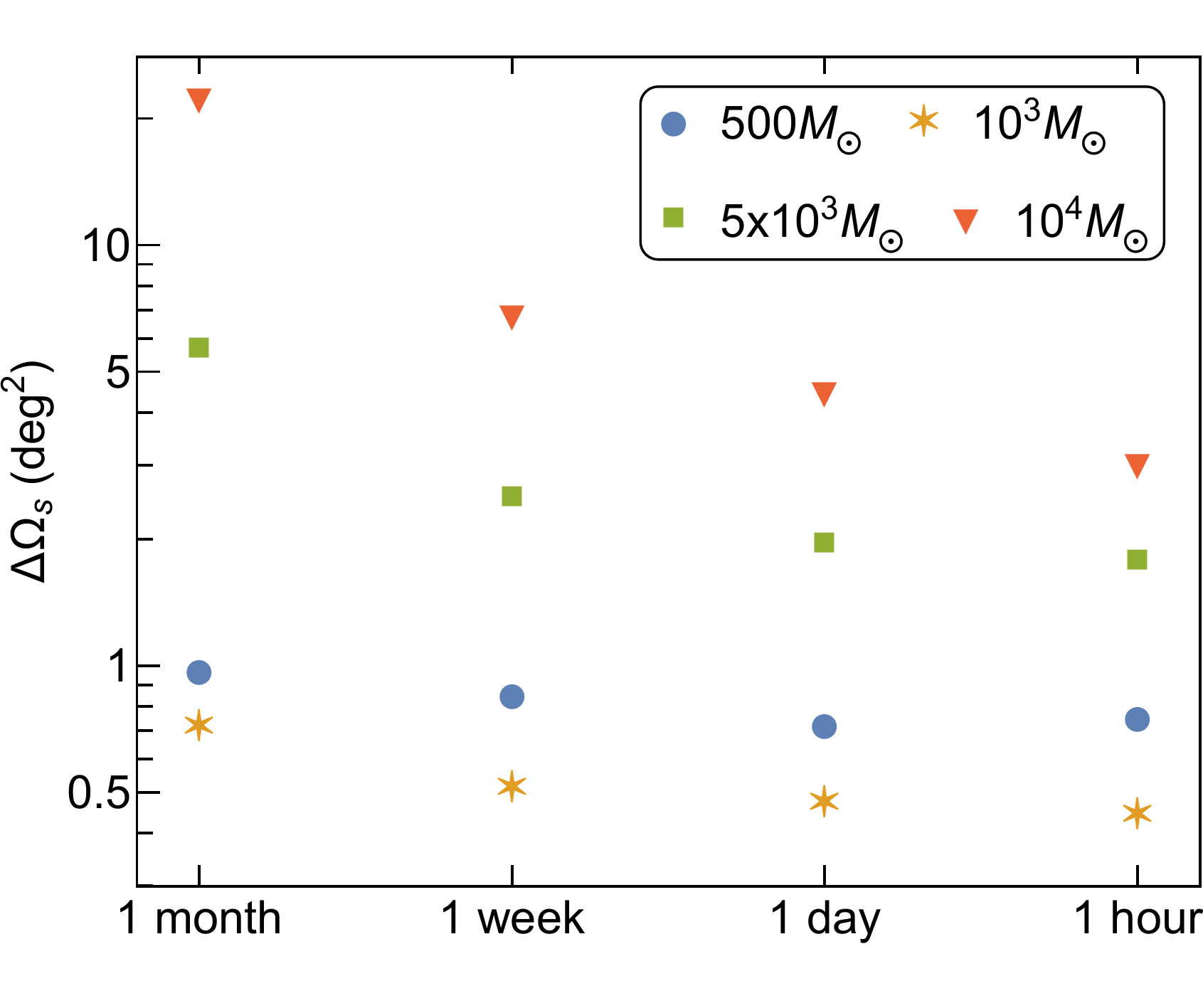}} 
   \end{subfigure} 
   \vspace{-0.35cm}
   \begin{subfigure}{\includegraphics[width=0.48\linewidth]{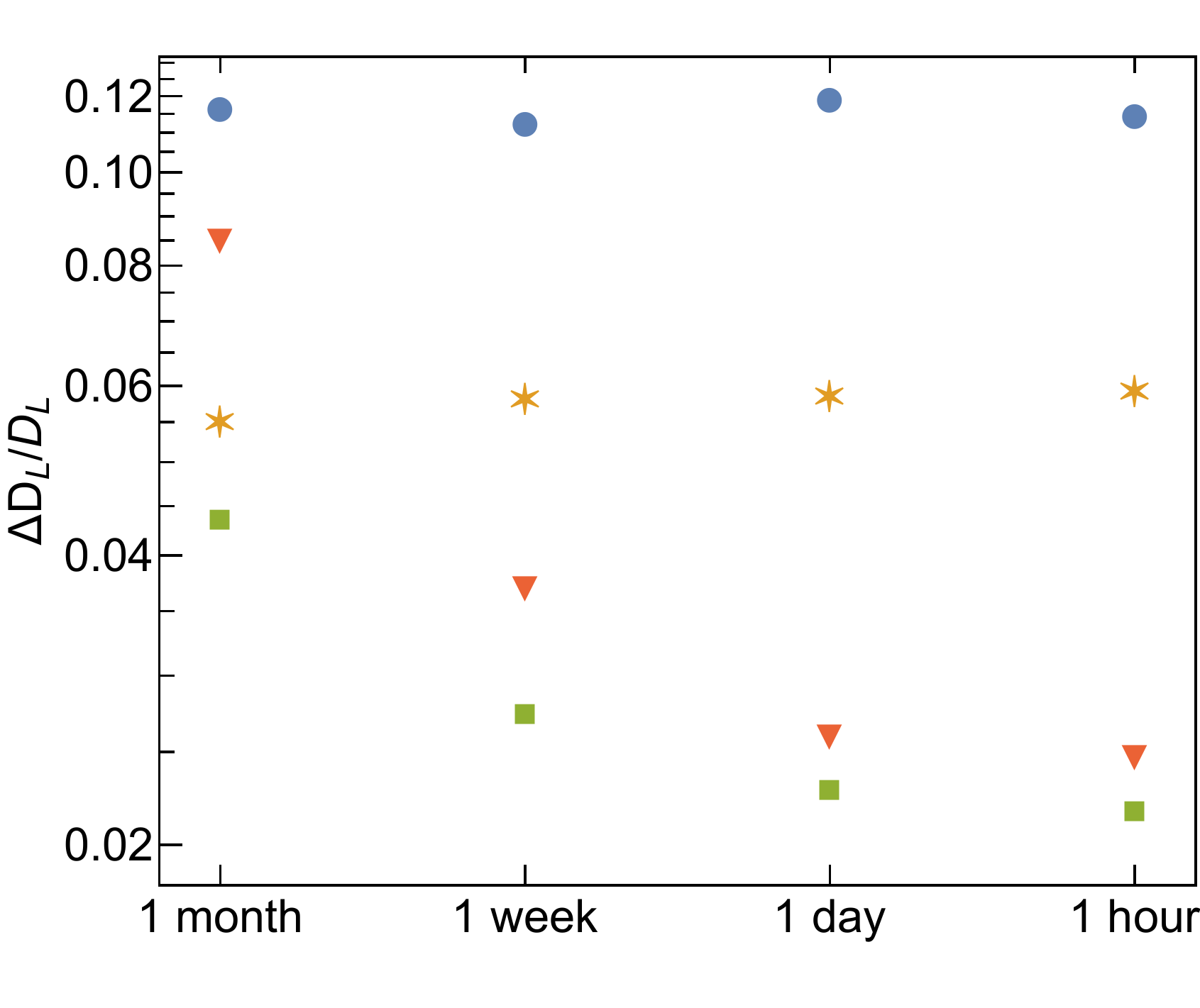}}
   \end{subfigure}
   \vspace{-0.35cm}
    \begin{subfigure}{\includegraphics[width=0.48\linewidth]{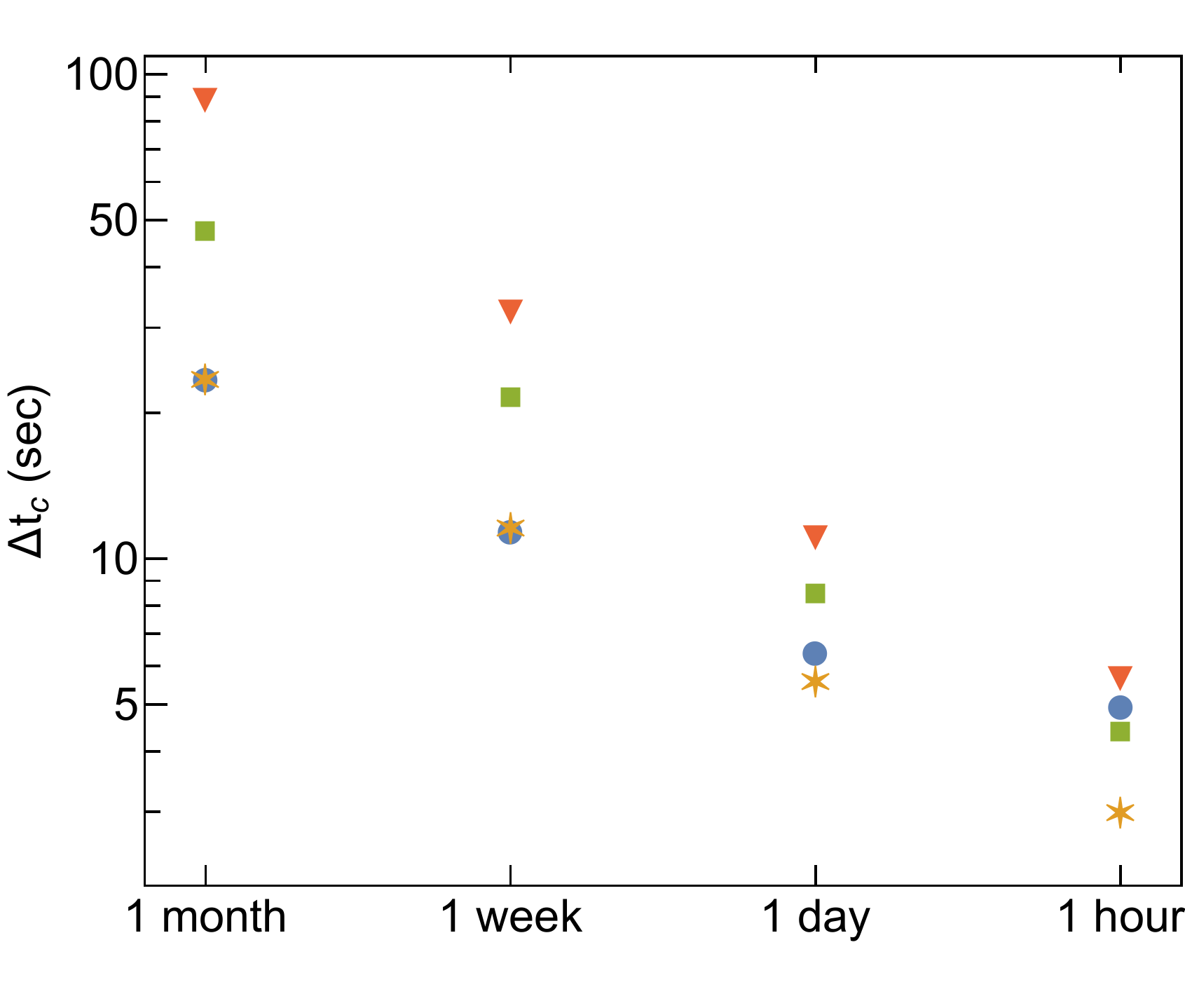}} 
    \end{subfigure}
    \begin{subfigure}{\includegraphics[width=0.48\linewidth]{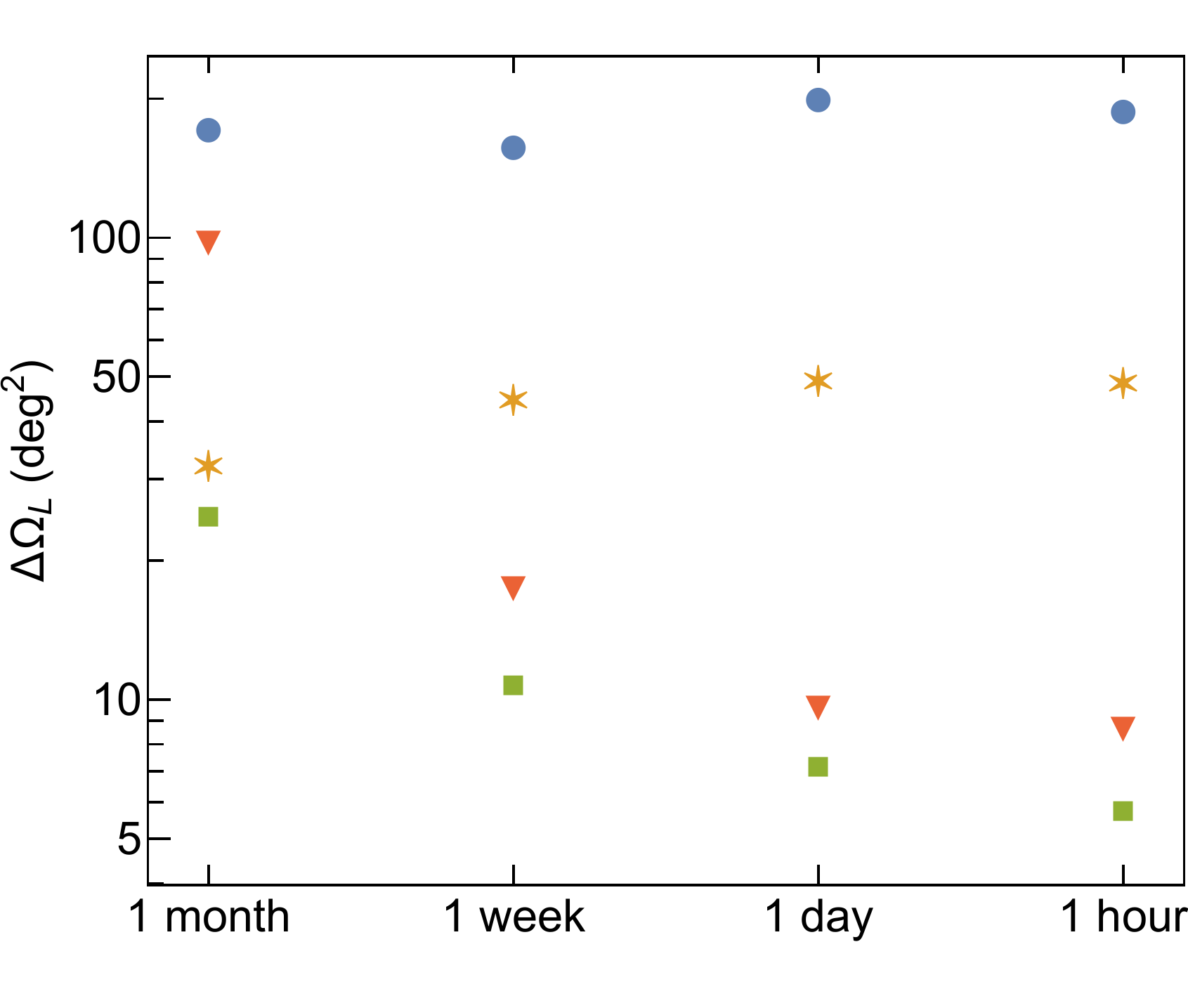}} 
    \end{subfigure}
     \caption{(Color online) Sky position, luminosity distance, coalescence time, and polarization resolution uncertainties as a function of time to coalescence for the same representative systems as considered in Fig.~\ref{snr}. The sky position and orientation of $10^3$ realizations of each system is generated in the same way as for Fig.~\ref{snr}. For each source, the mass ratio, dimensionless spin parameter, and luminosity distance are chosen to be the same as Fig.~\ref{snr}.}
      \label{extrinsic}
\end{figure*} 

\begin{figure*}[htp]
    \centering 
 \subfigure{\includegraphics[width=0.48\linewidth]{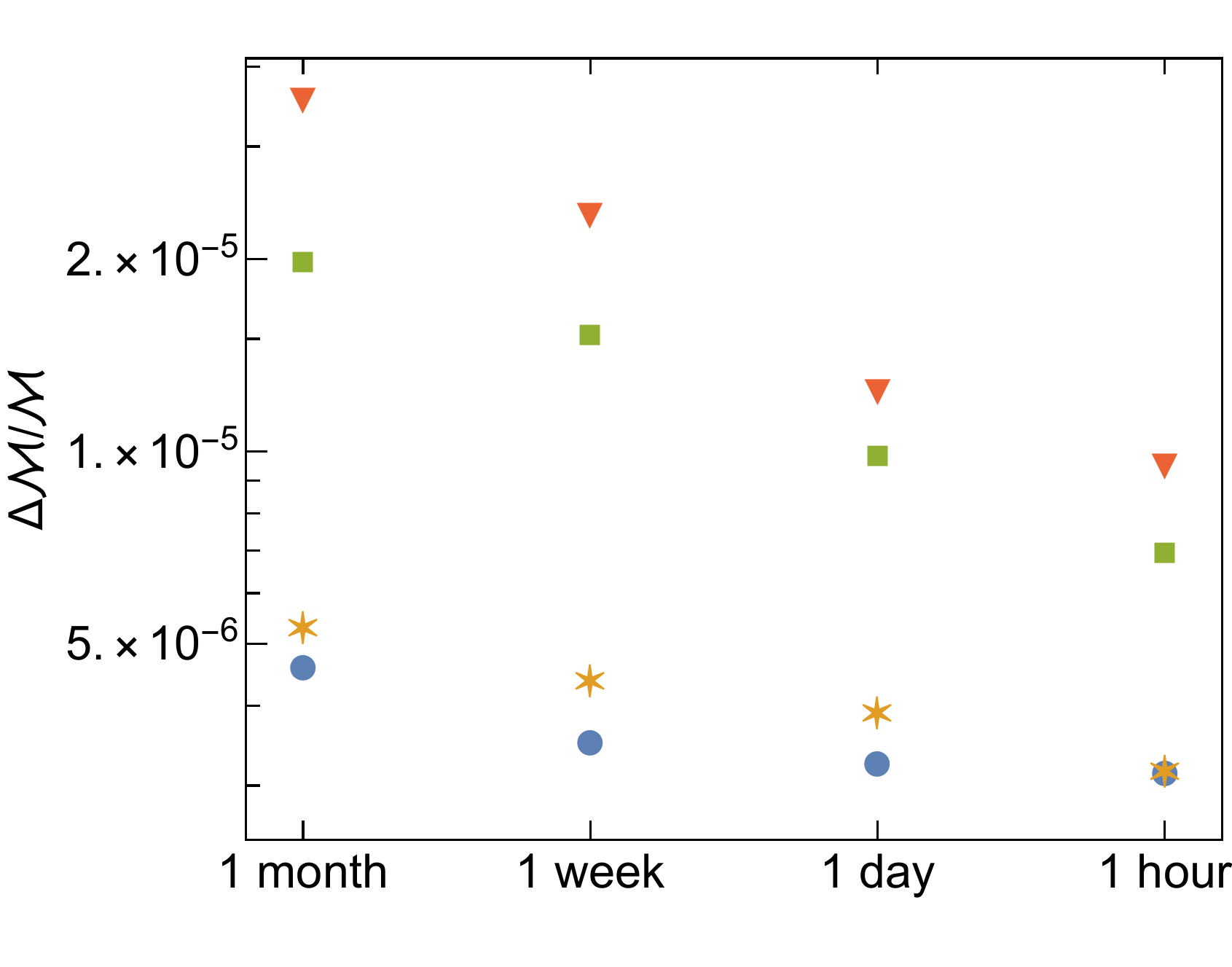}} 
    \subfigure{\includegraphics[width=0.48\linewidth]{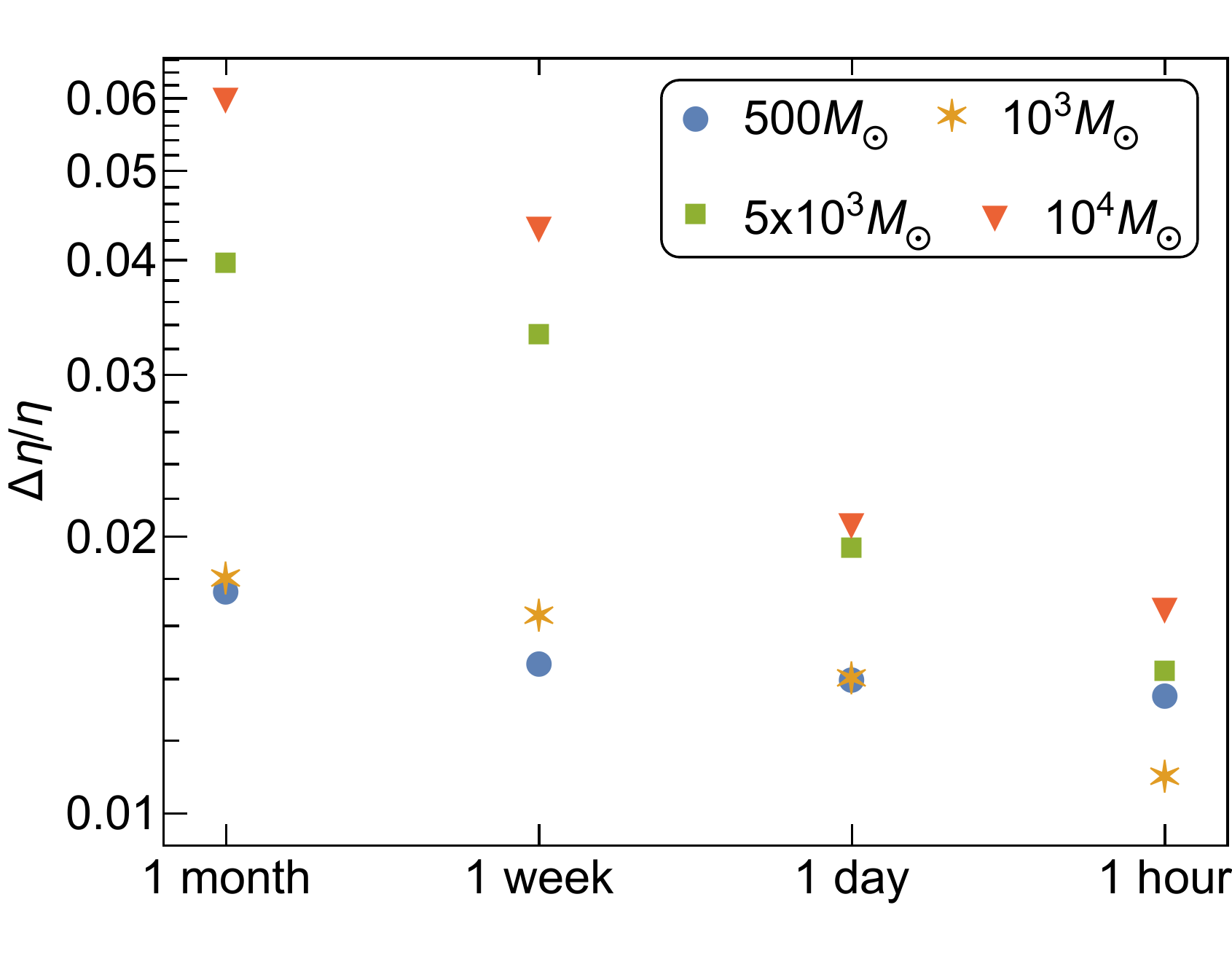}} \caption{Fractional uncertainties in chirp mass and symmetric mass ratio as a function of time to coalescence. All parameters are chosen to be the same as Fig.~\ref{snr}.}
     \label{intrinsic}
    \end{figure*}

We first discuss the signal-to-noise ratio of the events in the population. Besides the mass and distance, SNR depends on the position and orientation of the source in the sky. In our case, SNRs for all $10^3$ realizations corresponding to each representative system fall above LISA's SNR threshold $(\rho_{\rm th}=10)$. Moreover, following~\cite{Sesana:2010wy}, we impose an inversion accuracy condition for Fisher matrices $|\Gamma.\Sigma-\text{I}|\leq O(10^{-3})$, where $\Gamma$, $\Sigma$, and I are the Fisher matrix, the corresponding covariance matrix, and the identity matrix, respectively. Sources corresponding to Fisher matrices which do not satisfy the inversion accuracy condition are dropped out from our sample. For each system, more than $95\%$ of the sources pass the inversion accuracy condition except for the $500\, M_{\odot}$ system, for which $\sim 90\%$ of the sources (at 1 month before the coalescence) pass the conditions. For all the representative systems, the number of sources passing the inversion accuracy condition becomes more than $\sim 98\%$ before 1 hour of coalescence. Taking the Fisher matrices for those sources that pass the inversion accuracy condition, we calculate the 1$\sigma$ statistical errors. We plot the median values of these SNR and 1$\sigma$ errors as a function of time to coalescence. 

Figure \ref{snr} shows the median SNR as a function of time to coalescence for different total masses. For all systems, SNR increases as binaries evolve towards the merger. Since luminosity distance is fixed, lighter systems have lower value of SNR. As the binaries approach merger, SNR accumulates more rapidly for heavier mass systems compared to lighter ones. The system with $M=500\, M_{\odot}$ already gains a median SNR $\sim 20$ at 1 month before the merger. High mass system $10^{4}\, M_{\odot}$ has a median SNR of $\sim 35$ at 1 month before the merger, it accumulates most of the SNR during the last month of the merger. Median SNR for $10^{4}\,M_{\odot}$ becomes $\sim 250$ at 1 hour prior to the merger. In short, all the sources we consider have an SNR greater than $20$ at one week prior to the merger.

Figure~\ref{extrinsic} shows the time evolution of errors in the sky position $(\Delta\Omega_{s})$, errors in luminosity distance ($\Delta D_L/D_L$), time of coalescence ($\Delta t_c$) and polarization resolution ($\Delta\Omega_L$). The uncertainty in the sky position reduces as the binary approaches merger and accumulates SNR. Low mass sources ($500$, $10^3$)$\,M_{\odot}$ are better localized compared to the high mass sources ($5\times10^{3}$, $10^{4}$)$\,M_{\odot}$. This is because the low mass sources spend larger number of GW cycles in the LISA band as compared to the high mass sources, the modulation effects due to LISA's orbital motion around the Sun are able to break the degeneracies among the different parameters thus facilitating the better estimation of the sky localization. For high-mass systems, the uncertainty in sky position reduces rapidly during the last month before the merger as these systems accumulate most of their SNR during this period. The system with total mass $10^3\,M_{\odot}$ is the best localized source that has a median localization of $\sim 0.7 \,{\rm deg}^{2}$ even one month prior to the merger. This localization improves to $\sim 0.4\,{\rm deg}^{2}$ at one day before coalescence. For $10^{4}\,M_{\odot}$, accuracy on the sky position is $\sim 20\, {\rm deg}^{2}$ at one month prior to its merger. As the signal accumulates and SNR increases rapidly, this accuracy further improves to $\sim 5\,{\rm deg}^{2}$, one day before the merger and further to $\sim3{\rm\, deg}^{2}$ at one hour before merger. 

Instead of medians, it is also interesting to ask, for instance, the fraction of binaries in the simulated population that have angular resolution smaller than some representative number that has observational relevance. For $500\, M_{\odot}$ system, $25\%$ ($100\%$) of the binaries have localization errors smaller than the field of view (FOV) of Athena (LSST) one day before merger. The same for $10^3 \,M_{\odot}$ system is $40\%$ ($100\%$). For $5\times10^3\, M_{\odot}$ and $10^4\, M_{\odot}$ systems, the fractions of sources that fall within the FOV of Athena (LSST) at one day prior to merger are $6\%$ ($100\%$) and $8\%$ ($80\%$), respectively.

Next we turn our attention to errors in the luminosity distance. As expected the errors in $D_L$ depend strongly on the SNR of the source. As the SNR of the binary increases, uncertainty in the distance measurement reduces. Due to high SNR, high mass systems ($5\times10^3$, $10^4$)$\,M_{\odot}$ have better measurement of $D_L$ compared to low mass systems ($500$, $10^3$)$\,M_{\odot}$. For $500\, M_{\odot}$ and $10^3 \,M_{\odot}$ systems, the uncertainty in $D_{L}$ is almost constant i.e. $\sim 12\%$ and $6\%$ respectively, during the last month before merger since these systems do not gain much SNR during this period. The luminosity distance for ($5\times10^3$, $10^4$)$\,M_{\odot}$ systems can be measured with an accuracy of $\sim 4\%$ and $\sim 8\%$ respectively, at one month prior to merger. As the SNR for these systems increases, the uncertainty in $D_{L}$ reduces to $\sim2-3\%$ level, one day before the merger. Note that the errors quoted here are only due to the noise PSD of the detector. We do not take into account the systematic errors on $D_{L}$ measurement from phenomena like the weak-lensing effect and peculiar velocity, inclusion of which may deteriorate the errors in distance measurement \cite{wang2002universal,Holz:2004xx,bonvin2006fluctuations,shapiro2010delensing,tamanini2020peculiar,Mukherjee:2019qmm}. 

We also find that the time of coalescence $t_c$ can be measured to an accuracy of within $100$ sec for all the considered systems 1 month before coalescence. The uncertainty in the $t_{c}$ measurement reduces to within $10$ sec at 1 day before the merger, implications of which are discussed in the next section. In addition to these, we also compute the polarization resolution $\Delta\Omega_{L}$. For light systems ($500,10^3$)$\,M_{\odot}$ the accuracy on polarization resolution is as large as $200\,{\rm deg}^{2}$ and $50\,{\rm deg}^{2}$ respectively, 1 day before coalescence and for heavier systems ($5\times 10^{3},10^{4}$)$\,M_{\odot}$ it is around $7$ to $10\,{\rm deg}^{2}$ at one day before merger.
Trends in $\Delta \Omega_{\rm L}$ are opposite to that of $\Delta \Omega_{\rm S}$ but similar to those in the errors in the luminosity distance. This is due to the well-known degeneracy between $D_L$ and the inclination angle of the binary.

Further, though not the main focus of this paper, in Fig.~\ref{intrinsic}, we show the evolution of errors in chirp mass and symmetric mass ratio. Errors in the chirp mass and symmetric mass ratio reduce as the binary approaches merger. Chirp mass and symmetric mass ratio of low mass systems can be constrained better than high-mass binaries because of the large number of GW cycles that low mass sources have in the LISA band compared to high-mass binaries. For $500\, M_{\odot}$ and $10^3 \,M_{\odot}$ systems, chirp mass can be measured with a fractional accuracy of $\sim 10^{-6}$, 1 day before the merger. Fractional errors in the symmetric mass ratio for all the systems considered are around $\sim 1\mbox{--}2\%$ at 1 day before coalescence. 

The above-mentioned errors are calculated for a fixed $D_L$ and they will change as the $D_L$ varies. Owing to the degeneracies between $\mathcal{M}$, $D_L$ and angles $(\bar{\theta}_{S}, \bar{\phi}_{S}, \bar{\theta}_{L}, \bar{\phi}_{L})$, scaling of the errors with $D_L$ can be provided only partially. For a fixed mass, approximate scaling of SNR and errors with luminosity distance is expressed as: SNR $\approx D_L^{-1}$; $\Delta\Omega_{S}\approx D_{L}^{2}$ and $\Delta\Omega_{L}\approx D_{L}^{2}$; $\Delta D_L/D_L$, $\Delta \mathcal{M}/\mathcal{M}$, and $\Delta\eta/\eta$  $\approx D_L$. The actual errors may be slightly worse than the ones inferred from the partial scaling with $D_L$.

\section{Astrophysical Implications}\label{implications}
As we showed in the previous section, LISA can precisely measure the IMBBH parameters including the distance and sky location well in advance which can guide the follow-up observations using EM telescopes. We next discuss the important applications of these measurements for optimizing the observational strategies and detecting EM counterparts associated with IMBBH mergers.

\subsection{Optimization of observational strategies in the GW and EM bands}
A subset of the IMBBHs that lie on the lower mass side of the population may also be detectable by ground-based GW detectors, such as third-generation detectors Cosmic Explorer~\cite{Reitze:2019iox} and Einstein Telescope~\cite{LIGOScientific:2016wof,Maggiore:2019uih} as they merge at high frequencies which fall in the bandwidth of these detectors. Such observations in two different bands of the GW spectrum are usually referred to as multiband observations~\cite{Sesana:2016ljz,Vitale:2016rfr} and have very profound impact for fundamental physics and astrophysics~\cite{Liu:2020nwz} as several parameter degeneracies are lifted by the synergy of the two independent measurements~\cite{Gupta:2020lxa,Datta:2020vcj}.

It is evident from Fig.~\ref{extrinsic} that the LISA observations can help determine the arrival time of the GW signal in the CE band hours or days in advance with a precision that is about $\lesssim 10$ sec. This prior knowledge helps the detection of the signal in the CE band~\cite{Sesana:2016ljz, Pizzati:2021apa,Samajdar:2021egv}. Besides the detection, precise knowledge of the time of the merger in advance also helps the astronomy community to optimize their observational strategies which, in turn, depends on the mechanism they invoke for a potential EM transient. Prior knowledge of the merger time would help in deciding the time at which the EM telescopes should be pointed to the sky patch pinned down by LISA. Further, precise estimates of the component masses from the LISA signal could facilitate the assessment of detectability of the EM counterpart whose strength depends on the mass parameters in many contexts (see next subsection for specific examples). As several of the EM telescopes that can follow-up IMBBH mergers will have other key science objectives, this prior information from LISA can help in scheduling target of opportunity requests for telescope times.

\subsection{Prospects of detecting EM counterparts in optical and x-rays from IMBBHs}
Due to large uncertainties surrounding the exact mechanisms of EM emission around BBH mergers, here we will focus on two generic mechanisms which have been invoked in the context of stellar mass BBHs, one in the optical and the other in the x-rays. Both rely on accretion onto the remnant BH and/or onto the component BHs. 

\subsubsection{Optical flares from IMBBH mergers and their detectability  with LSST}
Consider an IMBBH merger in a gaseous environment such as in an AGN disc. Interaction of the merger remnant with the ambient gas-rich medium can lead to accretion onto the BH thereby producing electromagnetic flares. Here we closely follow a generic prescription discussed in \cite{McKernan:2019hqs} which has been used to interpret the optical flare observed by Zwicky Transient Facility (ZTF) in association with GW190521~\cite{graham2020candidate}. We summarize the important aspects of the method below and discuss the detectability of such optical flares by future optical survey facilities such as LSST.

The remnant BH formed by the merger receives a kick due to the loss of linear momentum through the anisotropic emission of gravitational waves during the last stages of coalescence~\cite{Peres:1962zz,fitchett1983influence,Favata:2004wz}. Hence, the newly born BH moves through the gaseous AGN disk displacing the bound gas along with it. As the bound gas interacts with the unperturbed gas outside, shocks are produced leading to bright hot spots in UV/optical bands. After a while, the BH gets out of the bound gas and directly interacts with the outside unperturbed gas leading to Bondi-Hoyle-Lyttleton (BHL) accretion as it is dragged by the gas. The corresponding Bolometric luminosity is given by~\cite{graham2020candidate}
\begin{eqnarray}
    L_{\rm bol}&\approx&2.5 \times 10^{47}{\rm erg}\, {\rm s}^{-1} \left( \frac{\eta_e}{0.1}\right)\left( \frac{M_{\rm rem}}{10^{3}\, M_{\odot}}\right)^{2} \nonumber \\
    &\times& \left(\frac{v_{k}}{200\, {\rm km}\, {\rm s}^{-1}} \right)^{-3} \left( \frac{\rho_m}{10^{-10}\, {\rm g}\, {\rm cm^{-3}}}\right)
    \label{eq:bhl},
\end{eqnarray}
where $\eta_e$ is the radiative efficiency, $M_{\rm rem}$ is the mass of the remnant BH, $v_k$ is the recoil kick velocity of the remnant and $\rho_m$ is the disk gas density. Since, the bolometric luminosity in the above equation is directly proportional to $M_{\rm rem}^{2}$ and $\rho_m$ and inversely proportional to $v_{k}^{3}$, the brightest emission will be for the modestly kicked ($v_{k} \sim {200\, {\rm km}\, {\rm s}^{-1}} $ in our case) remnant. 

We now consider the detectability of such flares with LSST.  Considering $30\,$sec exposure, LSST will reach a limiting (5$\sigma$) apparent magnitude of $m\sim24.5$ \cite{ivezic2019lsst}. With the prior information about the time of coalescence and angular resolution and assuming that the flare will at least last for several minutes, one can observe for an optimum time of $10$ minutes around the time of the merger. This leads to an improvement to the limiting magnitude of log$_{2.5}\sqrt{10\, \text{minutes}/30\, \text{sec}}\approx1.5$. We therefore set $m_{\rm LSST}=26$ as a fiducial detection limit \cite{tamanini2016science}.
The detection condition can finally be expressed as $m\leq m_{\rm LSST} $, where \cite{tamanini2016science,zombeck2006handbook} 
\begin{equation}
  m= \text{BC}+M_{\odot}^{\rm bol}+40-\frac{5}{2} \log_{10} \left( \frac{L_{\rm bol}}{L_{\odot}} \right)+5 \log_{10}\left(\frac{D_L}{\rm Gpc}\right) \, .
 	\label{eq:LSST_cond}
\end{equation}
Here, BC stands for Bolometric correction, $M_{\odot}^{\rm bol}$ is the solar Bolometric magnitude and $L_{\odot}$ is the Bolometric luminosity of the Sun. The values of these quantities are: $\text{BC}\approx1$ for LSST~\cite{tamanini2016science}, $M_{\odot}^{\rm bol} \approx 4.83$~\cite{williams2013sun}, and $L_{\odot} \approx 3.828 \times 10^{33}{\rm erg}\, {\rm s}^{-1}$~\cite{williams2013sun}. For $10^3 \,{M_\odot}$, assuming $M_{\rm rem}\approx 10^3\,{M_\odot}$ (ignoring mass loss to gravitational radiation), $\eta_e=0.1$, $\rho_m=10^{-10}\, {\rm g}\, {\rm cm^{-3}}$, the bolometric luminosity of the flare is $L_{\rm bol} \approx\, 2.5 \times 10^{47}\,{\rm erg}\,{s}^{-1}$, the corresponding apparent magnitude at $3$ Gpc is $m \approx 13.67$ which is far below the upper detection limit $m_{\rm LSST}$ and satisfies the detection condition. Hence, the source can be confidently detected by LSST. The sky resolution for $10^{3}\, M_{\odot}$ is $\sim0.5\,{\rm deg}^{2}$ a week prior to coalescence which is far below the FOV of LSST ($ \sim10\,{\rm deg}^{2}$). Moreover, $t_c$ is estimated with an accuracy of $\sim10$ sec for all the masses considered here. Therefore, LSST will have enough time to construct $ \approx 10^3$ point light curve of the object in the LISA error box and hence makes the detection of even any premerger optical counterpart possible.

Joint GW+EM detections of these mergers should give us valuable insights into the details of the environments in which IMBBHs merge. As the mass and the kick speed of the remnant can be inferred purely from GW observations, these joint detections can shed light on the density of the ambient medium as well as the efficiency of accretion, even in the absence of a detection, provided the telescope has sampled the sky patch sufficiently to detect any EM emission around its threshold. 

\subsubsection{X-ray emission from IMBBHs and detection prospects with Athena}
Next we discuss the detection of x-ray emission associated with IMBBH mergers. As there are no detailed models discussed in the literature for x-ray emission from IMBBH mergers, we again work with very general assumptions. If the IMBBH merger happens in dense environments, it is not unreasonable to assume that there can be accretion onto the component BHs or onto the remnant black holes (or, perhaps, both) which can emit in x-rays~\cite{Kaaret:2017tcn}. Detectors such as Athena should have the capability to search for such x-ray counterparts associated with IMBBH mergers. Without referring to any specific mechanisms, we consider the accretion-induced x-ray counterpart from IMBBHs and its detectability using Athena.

The best localized source in our analysis is $10^{3}\, M_{\odot}$ which has sky-position error $O(0.4\,{\rm deg}^{2})$, 1 day before the merger. Athena whose field of view is $\sim$ 0.4 ${\rm deg}^{2}$ thus gives us an exciting opportunity to observe the possible x-ray emission from the accreting IMBBHs. Besides this condition, for the detection of EM emission, the flux emitted by the source should be greater than the flux threshold of Athena. If both the BHs are accreting at a rate $f_{\rm Edd}$ and only a fraction ($10\%$) of emitted radiation is x-ray then the x-ray flux from the binary system is \cite{Caputo:2020irr}
\begin{equation}
    F_{\rm X} = 2 \times 10^{-13} f_{\rm Edd} \bigg(\frac{M}{M_{\odot}}\bigg)\bigg(\frac{\rm Mpc}{D_{L}}\bigg)^{2} {\rm erg\,cm^{-2}\,s^{-1}},
    \label{Xray Flux}
\end{equation}
where $M$ is the mass of the accreting BH and $D_L$ is the luminosity distance to the BH. The Eddington ratio ${\rm f_{Edd}}$ is the ratio that describes the fraction of the change in mass of the BH that arises from accretion, defined as  ${\dot M}_{\rm BH}/{\dot M}_{\rm Edd}$. As the growth of the BH is solely by accretion, we assume $f_{\rm Edd}=1$. In the above expression, it is also assumed that the accretion process has radiation efficiency $0.1$. The flux sensitivity for Athena for a 5$\sigma$ detection is \cite{McGee:2018qwb}

\begin{equation}
    F_{\rm Athena} = 10^{-15}\bigg(\frac{10^{3}\,{\rm s}}{\rm T_{int}}\bigg)^{1/2} {\rm erg\,cm^{-2}\,s^{-1}},
    \label{Athena flux}
\end{equation}
where ${\rm T_{int}}$ is the integration time for Athena. The minimum integration time required for a 5$\sigma$ detection of x-ray emission by Athena is given by equating Eqs.~\eqref{Xray Flux} and \eqref{Athena flux} and reads as

\begin{equation}
    {\rm T_{int}} \simeq 2\times 10^{-2} \frac{1}{f_{\rm Edd}^{2}} \bigg(\frac{D_{L}}{\rm Mpc}\bigg)^{4} \bigg(\frac{M_{\odot}}{M}\bigg)^{2} {\rm sec}\,.
\end{equation}
For $M=10^{3} \,M_{\odot}$ at $D_{L}=3$ Gpc, assuming the increase in mass of BH is almost completely accounted for by accretion ($f_{\rm Edd} \approx 1$) only, the minimum integration time for a $5\sigma$ detection is $\sim 18.75$ days. This integration time becomes $\sim5.5$ hours if the $D_L$ is reduced to $1$ Gpc. Hence a forewarning of a few hours by LISA prior to a possible x-ray counterpart associated with an IMBBH merger, at least in a subset of the sources, can lead to the detection of an EM counterpart by Athena. 
 
We conclude this section by noting that an independent redshift estimate from an EM counterpart can lead to cosmological parameter estimation~\cite{Schutz:1986gp}. Even in the absence of an EM counterpart detection, very precise localization can facilitate host galaxy identification using galaxy catalogs~\cite{MacLeod:2007jd,Gray:2019ksv,Chen:2017rfc,Mukherjee:2020hyn,LIGOScientific:2021aug}. Based on the luminosity distance and sky-position errors discussed earlier, if an independent estimate of redshift is available from either the galaxy surveys or EM counterparts, the Hubble-Lema$\hat{\text{\i}}$tre constant can be estimated to $\lesssim 10\%$ accuracy for a subset of the systems we consider here~\cite{Leandro:2021qlc,DES:2020nay}. Indeed, by the time LISA has launched the advanced ground-based detectors such as Voyager~\cite{mcclelland2015instrument} or third-generation GW detectors such as Einstein Telescope \cite{LIGOScientific:2016wof} or Cosmic Explorer \cite{Reitze:2019iox} may have achieved this level of accuracy~\cite{Borhanian:2020vyr}. Nevertheless, LISA observations should be very valuable as the two measurements come from GW observations at totally different frequency bands with complementary systematics. Consistency between the estimates of the two should significantly help in resolving the Hubble tension, the difference in the measured $H_0$ from supernova observations, and cosmic microwave background~\cite{DiValentino:2021izs}.

\section{Conclusions}\label{conclude}
We studied the projected parameter measurement uncertainties of intermediate mass binary black holes in the LISA band with a focus on their premerger localization and implications for EM followup campaigns. We found that in the best case scenario LISA has the potential to measure the errors in sky position $\sim 0.4\,{\rm deg^2}$ at one day prior to coalescence. These errors lie within the field of view (FOV) of EM telescopes such as Athena (FOV $\sim 0.4\,{\rm deg^2}$) and LSST (FOV $\sim 10\,{\rm deg^2}$). Moreover, LISA will be able to measure the luminosity distance within $\sim2\%$ (best case scenario) 1 day before merger, if the source is located at a luminosity distance of $3$ Gpc. These errors in the sky position and luminosity distance will roughly scale as $D_L^{2}$ and $D_L$, respectively. Furthermore, the time of coalescence for these binaries can be measured with errors $\lesssim 10$ sec an hour or days before their merger. The exciting possibility of locating these sources in the sky before their merger provides us a unique opportunity of exploring the environment of these binaries by possible EM followup, constraining the cosmological parameters such as Hubble-Lema$\hat{\text{\i}}$tre constant and exploring the formation of these binaries.

We end by stressing that we have used {\tt TaylorF2} waveform model for this analysis which does not account for higher order modes of gravitational waveforms~\cite{Arun:2004ff, Blanchet:2008je,Arun:2008kb,Garcia-Quiros:2020qpx,Pratten:2020ceb,Pan:2011gk,Nagar:2020pcj} as well as precessional effects induced by the misalignment of the spin vectors of the black holes with the orbital angular momentum of the binary~\cite{Kidder:1995zr,Arun:2008kb,Pan:2013rra,Pratten:2020ceb}. Both these effects lead to additional features in the waveforms as they introduce modulations in the phase and amplitude. Such features have been argued to significantly improve the localizability and distance estimation~\cite{Arun:2007hu, Trias:2007fp, Arun:2008kb,Stavridis:2009ys}. Hence, it is likely that the estimates quoted here may improve significantly upon inclusion of these features for a subset of IMBBHs which are asymmetric and have spin misalignment. A detailed study of these effects will be a topic of future work. We end by stressing that we use Fisher analysis as a way to assess the order of magnitudes of the errors associated with the measurement in each of the cases studied here. This method is expected to be reliable in the limit of high SNR \cite{Cutler_Flanagan,Balasubramanian:1995bm,vallisneri2008use}. In our analysis, the SNRs are always greater than $10$, however, detailed studies which numerically sample the likelihoods would be required to precisely quantify the measurement uncertainties.

\section*{Acknowledgments}
 We thank B. Sathyaprakash and A. Gupta for very useful discussions. We thank S. Datta for useful comments on the manuscript. K.G.A. and S.A.B acknowledge the  support from the Department of Science and Technology and Science and Engineering Research Board (SERB) of India via Swarnajayanti Fellowship Grant No. DST/SJF/PSA-01/2017-18. K.G.A also acknowledges support from SERB via Core Research Grant No. CRG/2021/004565, and MATRICS (Mathematical Research Impact Centric Support) Grant No. MTR/2020/000177. K.G.A, P.S.~and S.A.B. acknowledge partial support from the Infosys Foundation. 


\bibliographystyle{apsrev}
\bibliography{ref-list}
\end{document}